\documentclass[preprint,showkeys,amsmath,amssymb]{revtex4-1}
\usepackage{color,longtable}
\usepackage{graphicx}
\usepackage{amssymb,amsfonts,amsmath}
\usepackage{here}
\usepackage{dcolumn}
\usepackage{multirow}
\begin{document}

\title{Curing-induced filler aggregation in epoxy-amine systems}

\author{Yujiro Furuta}
\email{furuta@gel.t.u-tokyo.ac.jp}
\author{Rei Kurita}
\email{kurita@tmu.ac.jp}

\affiliation{%
\textit{Department of Physics, Tokyo Metropolitan University, 1-1 Minamioosawa, Hachiouji-shi, Tokyo, Japan.}
}%

\begin{abstract}
\noindent\textbf{Hypothesis:}
The macroscopic properties of polymer composites are governed by the dispersion and aggregation states of filler particles within a crosslinking matrix. 
Although curing transforms a liquid precursor into a solid network, its influence on filler aggregation remains insufficiently understood. 
We hypothesize that curing induces effective attractive interactions between filler particles, leading to aggregation even in non-Brownian systems.

\noindent\textbf{Experiment:}
Fluorescent polystyrene beads were homogeneously dispersed in a bisphenol F epoxy resin. 
Curing was initiated by adding trimethylhexamethylenediamine, and the evolution of the three-dimensional particle configurations was quantitatively examined using confocal laser fluorescence microscopy before and after completion of curing.

\noindent\textbf{Findings:}
Aggregation was enhanced during curing despite the absence of conventional attractive forces. 
The aggregation increment cannot be described solely by filler volume fraction but is governed by the mean interparticle gap $H$. 
Data collapse onto a linear scaling with the reduced gap parameter, identifying a geometric control parameter for curing-induced aggregation. 
This scaling demonstrates that curing dynamically generates an effective interaction whose spatial range scales with particle size, consistent with previously predicted rigidity-percolation-induced attractions. 
These findings establish a geometric criterion for predicting final dispersion states in curing polymer composites.
\end{abstract}


\maketitle


\section{Introduction}
Polymer composites consisting of crosslinked polymer matrices with dispersed filler particles are widely used in industrial applications such as paints, inks, adhesives, and flexible electronic devices~\cite{Saud2022-pc, Mendes-Felipe2019-iv, Aradhana2020-am, Li2023-xk}. 
In these materials, macroscopic properties including mechanical reinforcement, electrical conductivity, and multifunctionality are strongly governed by the dispersion and aggregation states of the fillers~\cite{Fu2008-pa, Oberdisse2006-xu, Wu2019-wz, Riviere2016-se, Aradhana2020-am, Stankovich2006-ml, Li2023-xk, Vysotskii1995-ps}. 
In practical fabrication processes, solidification is typically achieved through curing reactions that transform a liquid precursor into a crosslinked network~\cite{Abliz2013-cz, Jin2015-yt}. 
Because curing proceeds while fillers are suspended in the reacting medium, the evolving polymer network can modify the effective interactions between particles and thereby alter their spatial organization. 
Understanding how curing affects filler dispersion is therefore of both technological and fundamental importance.

Despite this importance, the physical mechanisms governing filler aggregation during curing remain poorly understood. 
Previous experimental studies have reported aggregation in curing epoxy systems~\cite{Shamurina1994-lk} and curing-dependent changes in electrical conductivity~\cite{Wu2019-wz}, suggesting structural reorganization of fillers. 
However, these studies did not directly resolve three-dimensional particle configurations nor identify the governing physical parameters. 
In particular, it remains unclear whether aggregation during curing originates from conventional interparticle forces, such as van der Waals or depletion interactions, or from mechanisms intrinsic to the evolving polymer network.

Recent simulation studies have proposed a distinct mechanism for curing-induced aggregation~\cite{Furuta2024-zt, Furuta2025-vu}. 
As crosslinking progresses, the polymer matrix undergoes rigidity percolation, forming a spatially heterogeneous elastic network. 
Near this percolation transition, fluctuations in local elastic constraints generate effective, elasticity-mediated attractive interactions between embedded filler particles. 
Importantly, the predicted interaction range is not determined by the polymer radius of gyration, as in classical depletion interactions~\cite{Asakura1954-gn, Asakura1958-sr}, but extends over distances comparable to the particle size. 
Moreover, the simulations indicated that both the spatial range and the interaction strength scale approximately with particle size. 
Such curing-induced interactions therefore differ fundamentally from conventional short-range polymer-mediated attractions and instead arise dynamically from the evolving elastic network.

Although the rigidity-percolation-mediated attraction proposed in simulations provides a plausible mechanism for curing-induced aggregation~\cite{Furuta2024-zt, Furuta2025-vu}, direct experimental verification remains lacking. 
The simulations primarily captured the evolution of network connectivity during curing, whereas the influence of additional constraints, such as angular rigidity and post-bond rearrangements, on the resulting effective interactions has not been fully established. 
It therefore remains unclear whether such elasticity-mediated interactions emerge in real epoxy-amine systems and whether their predicted particle-size-dependent range and strength can be observed experimentally. 
Furthermore, the governing geometric parameters that control the final aggregation state under practical curing conditions have yet to be systematically identified.

To address these issues, we experimentally investigate filler aggregation during curing as a function of particle size and volume fraction. 
Fluorescent polystyrene beads are employed as model fillers, enabling direct three-dimensional visualization of particle configurations before and after network formation using confocal laser fluorescence microscopy. 
We demonstrate that aggregation is enhanced during curing even in non-Brownian systems where conventional attractive forces are insufficient to induce contact within the relevant time scale. 
Furthermore, we show that the aggregation increment is governed not by volume fraction alone, but by a dimensionless geometric parameter defined by the mean interparticle gap normalized by particle size. 
These results provide experimental evidence consistent with elasticity-mediated interactions emerging during curing and establish a predictive geometric criterion for controlling final dispersion states in polymer composite materials.

\section{Materials and Methods}

\subsection{Preparation of samples}
A total of $4.0 \times 10^2$ mg of bisphenol F epoxy resin (EPICLON 830, DIC Corporation; equivalent weight per epoxide group: 165--177 g/mol) was placed in an Eppendorf tube, and 32 mg of dry fluorescent polystyrene beads (Fluoro-Max, diameters $a = 5$ and $8~\mathrm{\mu m}$ with polydispersity 12\%, Thermo Fisher Scientific; hereafter FPS) were added.
The suspension was manually mixed with a spatula until no visible agglomerates remained, followed by homogenization using a planetary centrifugal mixer (THINKYMIXER, THINKY Corporation) for 2 min and a subsequent 3 min defoaming cycle. 
Trimethylhexamethylenediamine (90 mg; mixture of the 2,2,4- and 2,4,4-isomers; Tokyo Chemical Industry Co., Ltd.) was then added, and the mixture was stirred manually.
The compositions were adjusted to obtain the desired filler volume fraction $\varphi$.
The epoxy and amine components were mixed in stoichiometric proportions to ensure complete crosslinking.
Approximately 0.2 g of the resulting epoxy--amine--FPS dispersion was injected into a sample cell and sealed with a cover glass using silicone grease. 

\subsection{Evaluation of aggregation fraction}
We quantified the curing ratio using Fourier transform infrared spectroscopy (FT-IR; FT/IR-4600, JASCO Corporation, equipped with a DuraSamplIR II accessory).
The characteristic absorption peak at 915 cm$^{-1}$, assigned to the epoxy group, progressively decreased, while the peak at 1115 cm$^{-1}$, corresponding to the C–O stretching of aliphatic secondary alcohols\cite{Infrared-Spectroscopy-Atlas-Working-Committee-ed-Brezinski1991-qk} formed during curing, increased in intensity (see Fig. S1).
The curing ratio, $\alpha$, was calculated according to the following equation:
\[
\alpha = 1 - \frac{I_{\rm{epoxy}}/I_{\rm{ether}}}{I^{0}_{\rm{epoxy}}/I^{0}_{\rm{ether}}}.
\]
Here, $I_{\rm{epoxy}}$ and $I_{\rm{ether}}$ denote the intensities of the epoxy and aromatic ether peaks, respectively, while $I^{0}_{\rm{epoxy}}$ and $I^{0}_{\rm{ether}}$ represent their corresponding initial intensities.
Normalization with respect to the aromatic ether peak was employed to minimize the influence of noise.
Using this approach, we measured the temporal evolution of the curing ratio $\alpha$, confirming that no further change in $\alpha$ was observed over a period of seven days.

Viscosity measurements of the epoxy–amine mixture during curing were performed using a rheometer (MCR 302e, Anton Paar) equipped with a cone–plate geometry. 
All measurements were conducted under steady shear at 23 ${}^\circ$C.
The shear rate was varied logarithmically over the range 0.01–1000 s$^{-1}$, with sixteen data points collected during both ascending and descending sweeps (32 data points in total). 
The results demonstrated negligible shear-rate dependence within the investigated range, confirming Newtonian behavior.
The representative viscosity at each curing time was obtained by averaging the 20 data points (ascending and descending sweeps) measured within the 0.01–10 s$^{-1}$ range. 
Measurements were carried out at 0, 0.5, 1, and 2 h after mixing to characterize the viscosity evolution accompanying the progression of the curing reaction.

Samples were observed using confocal laser scanning fluorescence microscopy (FV-1000D, IX-81, Olympus) at the initial stage and after completion of curing. 
Observations at the early stage were performed 5 min after sealing the sample cell to allow relaxation of flow induced during sample preparation, whereas observations after curing were conducted at least 7 days later. 
A square region of $640~\mathrm{\mu m} \times 640~\mathrm{\mu m}$ was imaged in the xy-plane. Along the z-axis, optical slices were acquired at intervals of $1~\mathrm{\mu m}$ over a total thickness of $20~\mathrm{\mu m}$, enabling three-dimensional reconstruction of particle positions.
Observations were performed at seven distinct locations within a sample and subsequently averaged.
Due to instrumental time constraints, data acquisition was performed at discrete time points rather than continuously; consequently, the measurement positions were not precisely identical.

Particle centers were identified using a custom Python-based image analysis code. After Gaussian smoothing of the fluorescence images (smoothing width $\sigma_g = 1.2$ pixels for $a = 5~\mathrm{\mu m}$ and $\sigma_g = 1.5$ pixels for $a = 8~\mathrm{\mu m}$), local intensity maxima were detected and assigned as particle centers.
Two particles $i$ and $j$ were regarded as aggregated when their center-to-center distance $d_{ij}$ satisfied $d_{ij} \le 1.12 a$.
The threshold factor 1.12 accounts for the manufacturer-specified 12\% size tolerance of the beads. 
Even when the threshold is set at 17\%, the results remain essentially unchanged.
The aggregation fraction $\Psi_{\mathrm{agg}}$ was defined as the fraction of particles having at least one neighboring particle satisfying the above criterion relative to the total number of fillers in the observation volume. 
Note that $\Psi_{\mathrm{agg}}$ is defined on a particle basis rather than a cluster basis; thus, a dimer and a trimer contribute two and three aggregated particles, respectively.
Thus, $\Psi_{\mathrm{agg}}$ serves as a quantitative measure of filler aggregation.

\section{\label{sec:level3}Results}
\subsection{\label{sec:level3A}Curing kinetics and viscosity evolution}
\begin{figure}[htbp]
\begin{center}
\includegraphics[width=150mm]{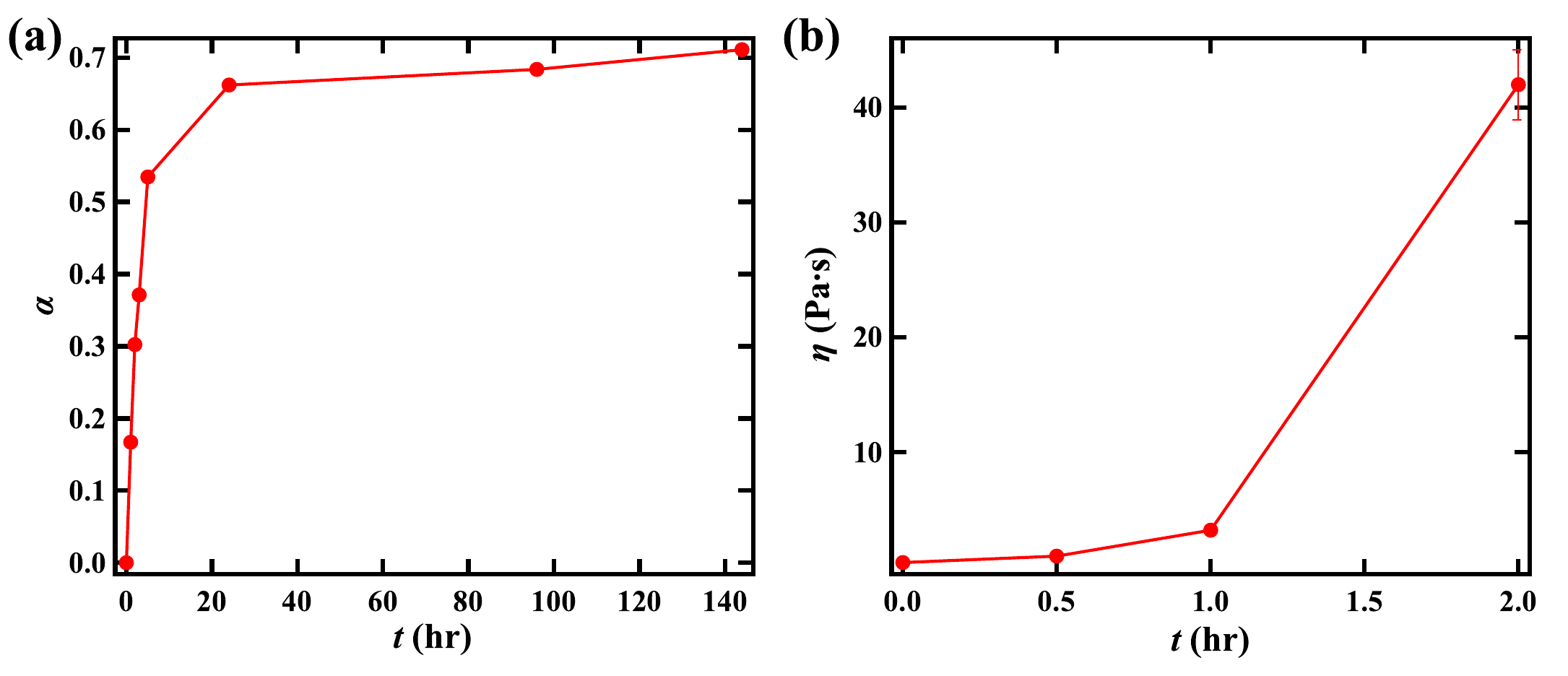}
\caption{\label{fig:FTIR} 
(a) Temporal evolution of the curing ratio, $\alpha$, determined from the peak intensity ratios in the FT-IR spectra. Measurements were conducted at $t = 0, 1, 2, 3, 6, 24, 96,$ and $144$ h.
(b) Temporal evolution of viscosity $\eta$. The viscosity increases to approximately 40 Pa$\cdot$s (nearly two orders of magnitude higher than the initial value) within about 2 h, after which it rises sharply, rendering further measurements difficult.
}
\end{center}
\end{figure}

We first present the temporal evolution of the curing process.
Fig.~\ref{fig:FTIR}(a) illustrates the time dependence of the degree of cure, $\alpha$, determined from the peak intensity ratios in the FT-IR spectra.
The value of $\alpha$ increases rapidly and approaches saturation at approximately $t = 24$ h, indicating that the curing reaction is essentially complete after seven days.
Concurrently, the viscosity $\eta$ increases with the progression of curing.
Fig.~\ref{fig:FTIR}(b) shows the temporal evolution of $\eta$, demonstrating that the viscosity rises by nearly two orders of magnitude within a few hours.
Because the mobility of the fillers is inversely proportional to the matrix viscosity, filler aggregation predominantly occurs within the first two hours, during which the viscosity undergoes a pronounced increase.

\subsection{\label{sec:level3B}Influence of filler size and volume fraction on aggregation during curing}
\begin{figure}[htbp]
\begin{center}
\includegraphics[width=150mm]{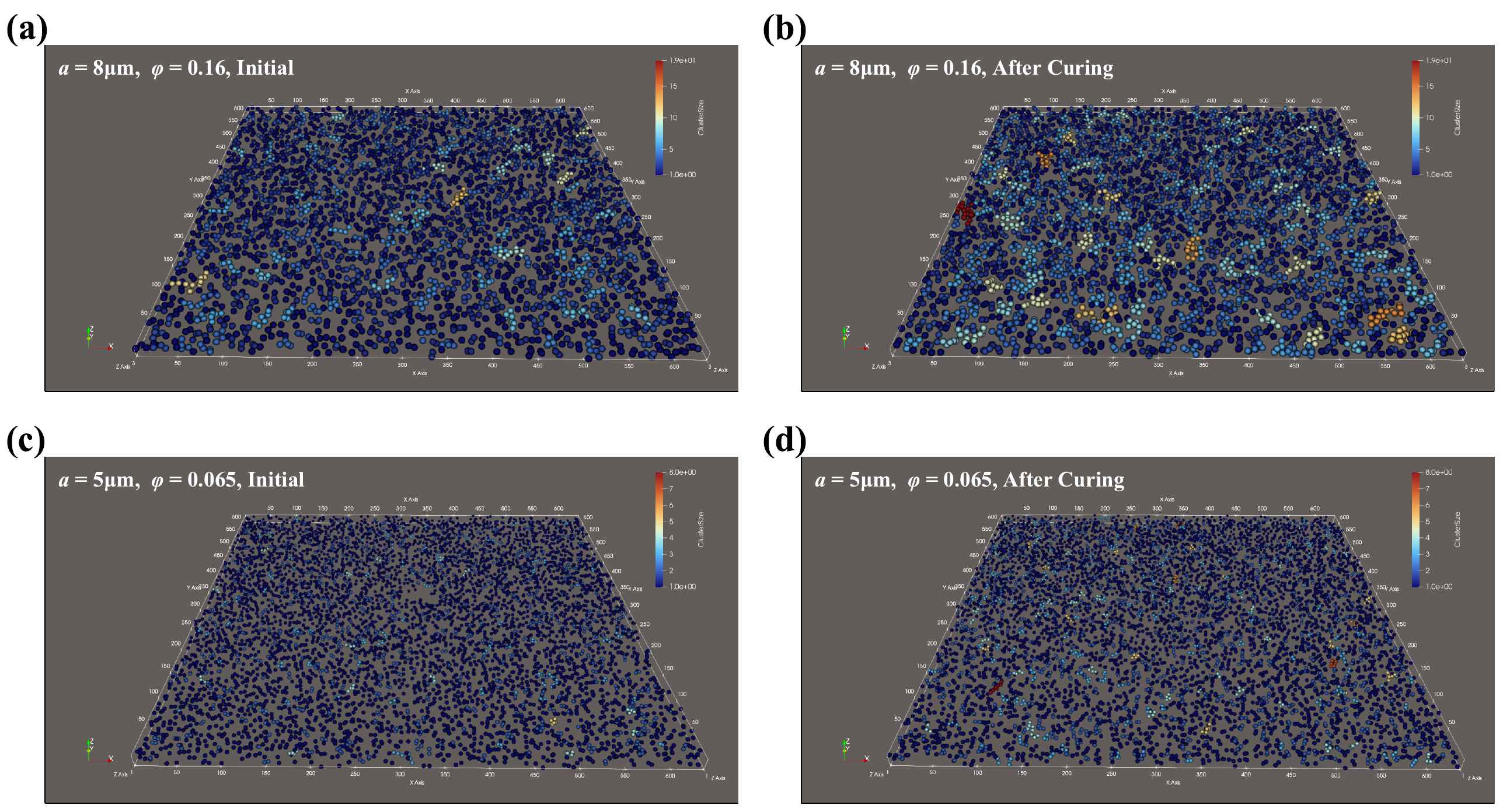}
\caption{\label{fig:macro} 
Three-dimensional reconstructions from confocal fluorescence microscopy images at the initial curing stage (a) and after curing (b) for $a = 8\,\rm{\mu m}$ and $\varphi = 0.16$. 
Three-dimensional filler configurations at the initial curing stage (c) and after curing (d) for $a = 5\,\rm{\mu m}$ and $\varphi = 0.065$. 
The particle color represents the cluster size. ParaView was employed for three-dimensional reconstruction.}
\end{center}
\end{figure}

To visualize the structural changes induced during curing, we first examine representative filler particle configurations reconstructed from confocal fluorescence images.
Fig.~\ref{fig:macro}(a) and (b) show the particle configurations at the early stage of curing and after completion of curing, respectively, for $a = 8\,\rm{\mu m}$ and $\varphi = 0.16$. 
The color scale represents the cluster size $s$ to which each particle belongs. 
Dark blue ($s = 1$) corresponds to isolated particles. 
A qualitative increase in both cluster number and cluster size is observed when comparing the early and post-curing states. 
Furthermore, the aggregated clusters exhibit anisotropic morphologies in both stages. 
Such anisotropy is likely attributable to hydrodynamic interactions commonly observed in colloidal dispersion systems~\cite{Furukawa2010-ag}. 
We note that the system does not form system-spanning aggregates~\cite{Kohjiya2008-yf}; instead, the aggregates remain spatially dispersed as isolated clusters throughout the observation region.
A similar trend was observed for particles with $a = 5 ,\rm{\mu m}$ at $\varphi = 0.065$, as well as under other combinations of particle size and volume fraction (Fig.~\ref{fig:macro}(c) and (d)).

\begin{figure}[htbp] 
\begin{center} 
\includegraphics[width=150mm]{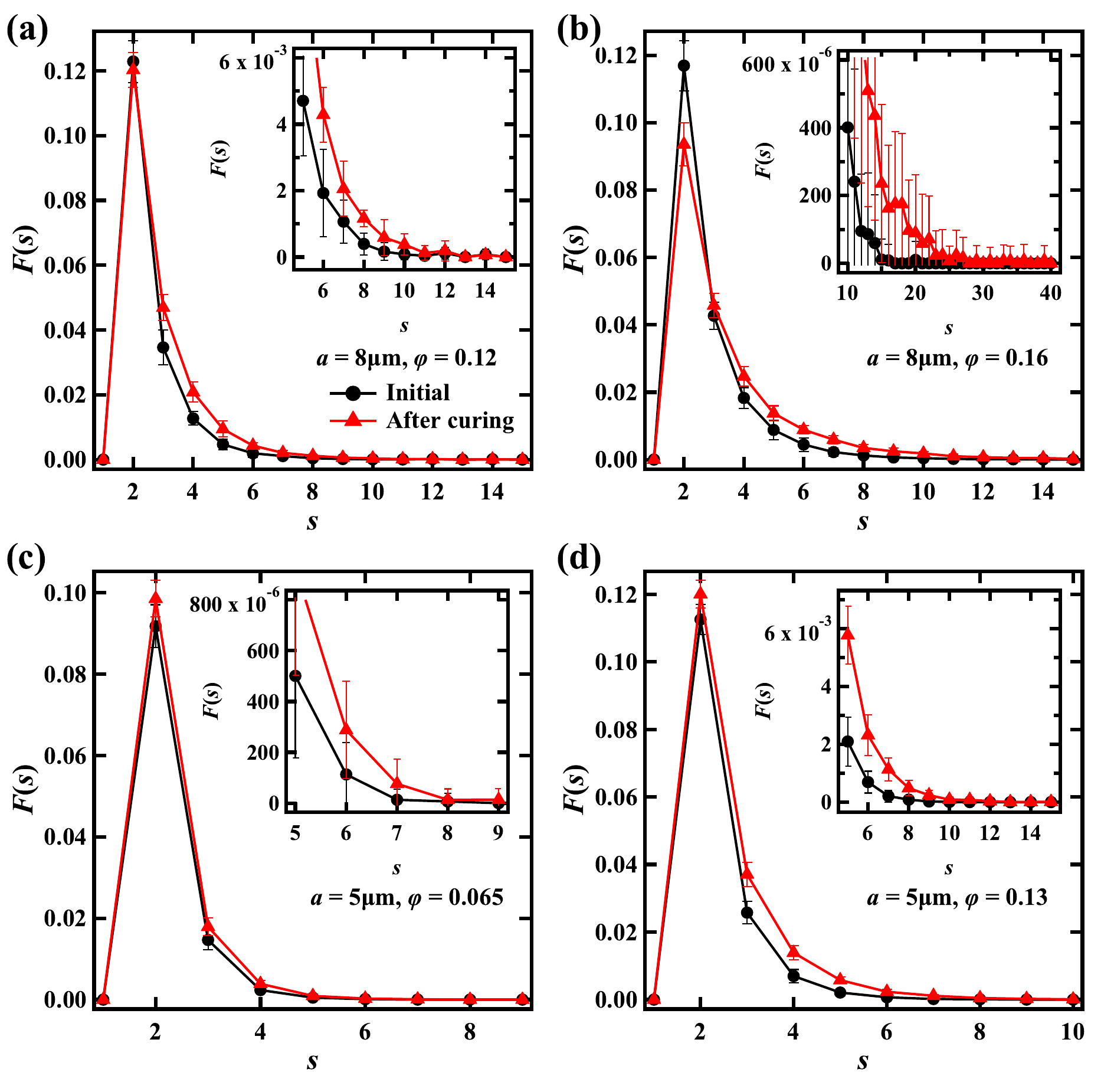}
\caption{\label{fig:size_dist} 
Cluster size distribution $F(s)$ at the initial curing stage and after curing under each experimental condition: 
(a) $a = 8\,\rm{\mu m}$, $\varphi = 0.12$; 
(b) $a = 8\,\rm{\mu m}$, $\varphi = 0.16$; 
(c) $a = 5\,\rm{\mu m}$, $\varphi = 0.065$; 
(d) $a = 5\,\rm{\mu m}$, $\varphi = 0.13$. 
The black line with filled circles represents the distribution at the onset of curing, whereas the red line with filled triangles denotes the distribution after curing.
Inset: Enlarged views of the large $s$ region in the $F(s)$ distributions for each experimental condition.}
\end{center} 
\end{figure}

To quantify these structural changes, we next analyze aggregation involving multiple filler particles. 
Fig.~\ref{fig:size_dist} shows the cluster size distribution $F(s)$ within the observation region, defined as $F(s) = \langle n(s)/n_{\mathrm{all}} \rangle$, where $n(s)$ denotes the number of clusters of size $s$, $n_{\mathrm{all}}$ is the total number of filler particles within the filed of view, and $\langle \cdots \rangle$ represents ensemble averaging over independent observation regions.
Fig.~\ref{fig:size_dist}(a) presents $F(s)$ for $a = 8\,\rm{\mu m}$ and $\varphi = 0.12$. 
Under these conditions, the fraction of dimers ($s = 2$) remains essentially unchanged, whereas the fraction of larger clusters ($s \geq 3$) increases.
As the volume fraction is increased to $\varphi = 0.16$, the fraction of dimers decreases, and the formation of larger clusters becomes progressively more pronounced (Fig.~\ref{fig:size_dist}(b)).

In contrast, Fig.~\ref{fig:macro}(c) and (d) show the spatial distributions for $a = 5\,\rm{\mu m}$ and $\varphi = 0.065$ before and after curing, respectively. 
The corresponding cluster size distributions are presented in Fig.~\ref{fig:size_dist}(c) and (d). 
Compared with the $a = 8\,\rm{\mu m}$ case, the aggregation-promoting effect is weaker: the increase after curing is mainly observed in relatively small clusters, whereas the formation of larger aggregates is less pronounced.

\subsection{\label{sec:level3C}Relationship between aggregation-promoting effect and volume fraction}

To quantitatively assess the aggregation-promoting effect of the curing process, we evaluated the aggregation fraction $\Psi_{\rm agg}$ at the early stage of curing and after completion of curing. 
In our system, the initial viscosity is relatively high (approximately $0.4~\mathrm{Pa \cdot s}$), and the filler particle diameters are large ($a = 5$ and $8~\mathrm{\mu m}$), such that the particles behave as non-Brownian. 
In the absence of any additional attractive interaction emerging during curing, the particle configuration would therefore be expected to remain unchanged, and the change in aggregation fraction should be negligible.

\begin{figure}[htbp]
\includegraphics[width=150mm]{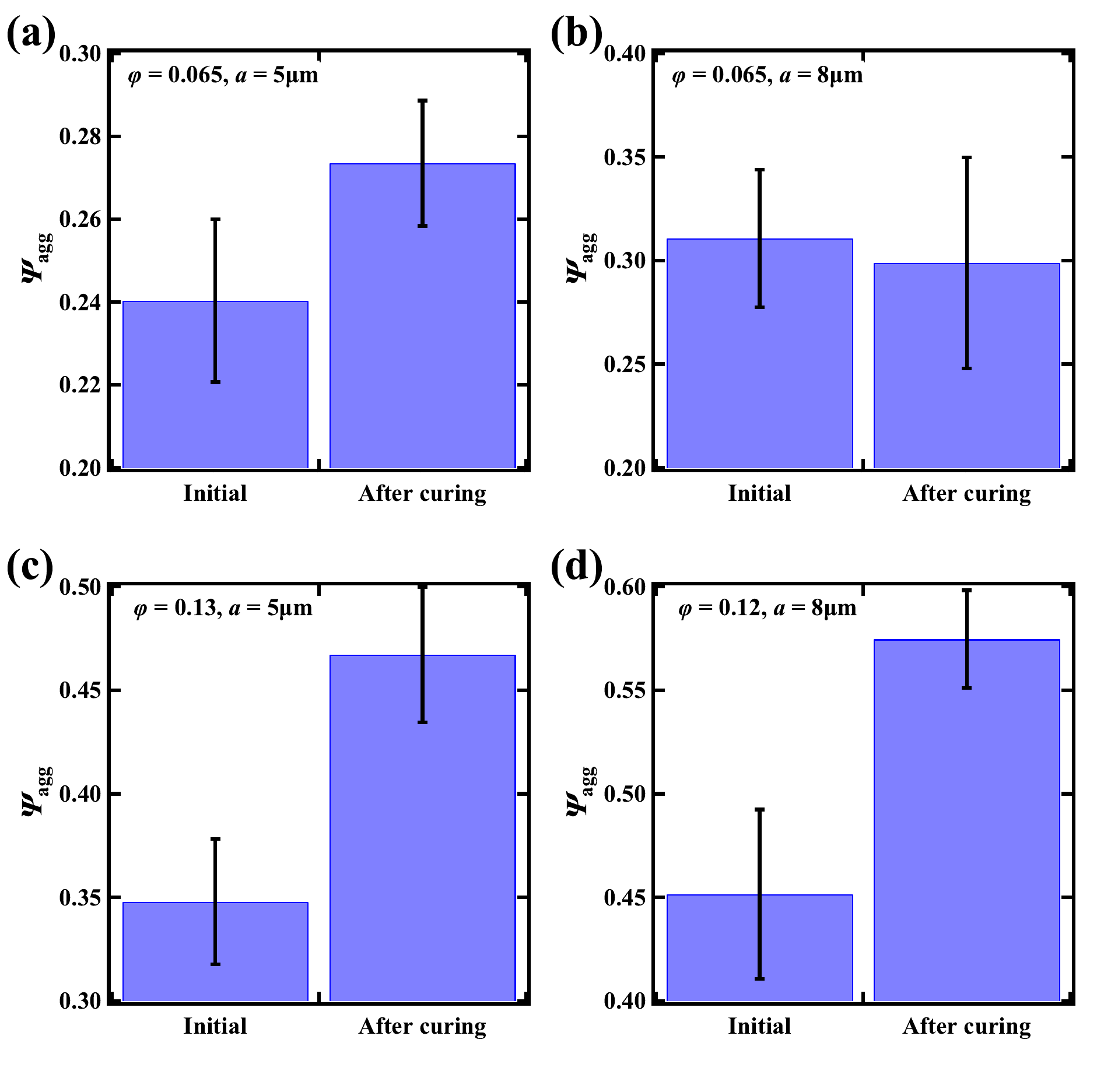}
\caption{\label{fig:psi}
Aggregation fraction, $\Psi_{\rm agg}$, at the early and final stages of curing. 
Panels (a) and (b) correspond to $\varphi = 0.065$ with particle diameters $a = 5\,\rm{\mu m}$ and $8\,\rm{\mu m}$, respectively. 
Panels (c) and (d) correspond to $a = 5\,\rm{\mu m}$ at $\varphi = 0.12$ and $a = 8\,\rm{\mu m}$ at $\varphi = 0.13$, respectively. 
Error bars represent standard deviations obtained from independent measurements.
}
\end{figure}

Fig.~\ref{fig:psi}(a) and (b) show the variation in $\Psi_{\rm agg}$ before and after curing for particles with diameters $a = 5$ and $8\,\rm{\mu m}$ at the same volume fraction $\varphi = 0.065$. 
Although the volume fraction is identical, $\Psi_{\rm agg}$ differs markedly depending on particle size. 
For $a = 5\,\rm{\mu m}$, $\Psi_{\rm agg}$ increases after curing. 
Although the error bars overlap, statistical analysis yielded a p-value of $2.64 \times 10^{-7}$, confirming that the increase is statistically significant. 
In contrast, for $a = 8\,\rm{\mu m}$ at $\varphi = 0.065$, no significant difference is observed, indicating that curing does not promote appreciable aggregation under this condition.

At higher volume fractions, however, a clear enhancement of aggregation is observed. 
Around $\varphi = 0.12$, $\Psi_{\rm agg}$ increases noticeably after curing (Fig.~\ref{fig:psi}(c) and (d)), and the increase becomes more pronounced at $\varphi = 0.16$. 
These results indicate that while the volume fraction is an important parameter, it does not uniquely determine the aggregation-promoting effect.

\subsection{\label{sec:level3D}Relationship between aggregation-promoting effect and interparticle gap}

We now focus on the mean interparticle gap $H$, noting that even at the same volume fraction, $H$ differs depending on the filler particle size. 
In classical depletion interactions as well as in the curing simulations, the surface-to-surface separation plays a central role in determining the effective interaction potential~\cite{Furuta2024-zt, Furuta2025-vu, Asakura1954-gn, Asakura1958-sr}. 
The mean interparticle gap is given by
\[
H = a \left( \sqrt[3]{\frac{\pi}{6\varphi}} - 1 \right).
\]
This relation implies that even at identical volume fractions, the effective interaction experienced by particles can differ substantially depending on $a$. 
This expectation is consistent with the observation that $\Psi_{\rm agg}$ is smaller for larger particle sizes at fixed $\varphi$.

To quantitatively evaluate the relation between $H$ and the aggregation-promoting effect, we define the aggregation increment as
\[
\Delta \Psi_{\rm agg} = \Psi_{\rm agg}^{\rm end} - \Psi_{\rm agg}^{\rm ini},
\]
where $\Psi_{\rm agg}^{\rm end}$ and $\Psi_{\rm agg}^{\rm ini}$ denote the aggregation fractions after and before curing, respectively.

\begin{figure}[htbp]
\includegraphics[width=150mm]{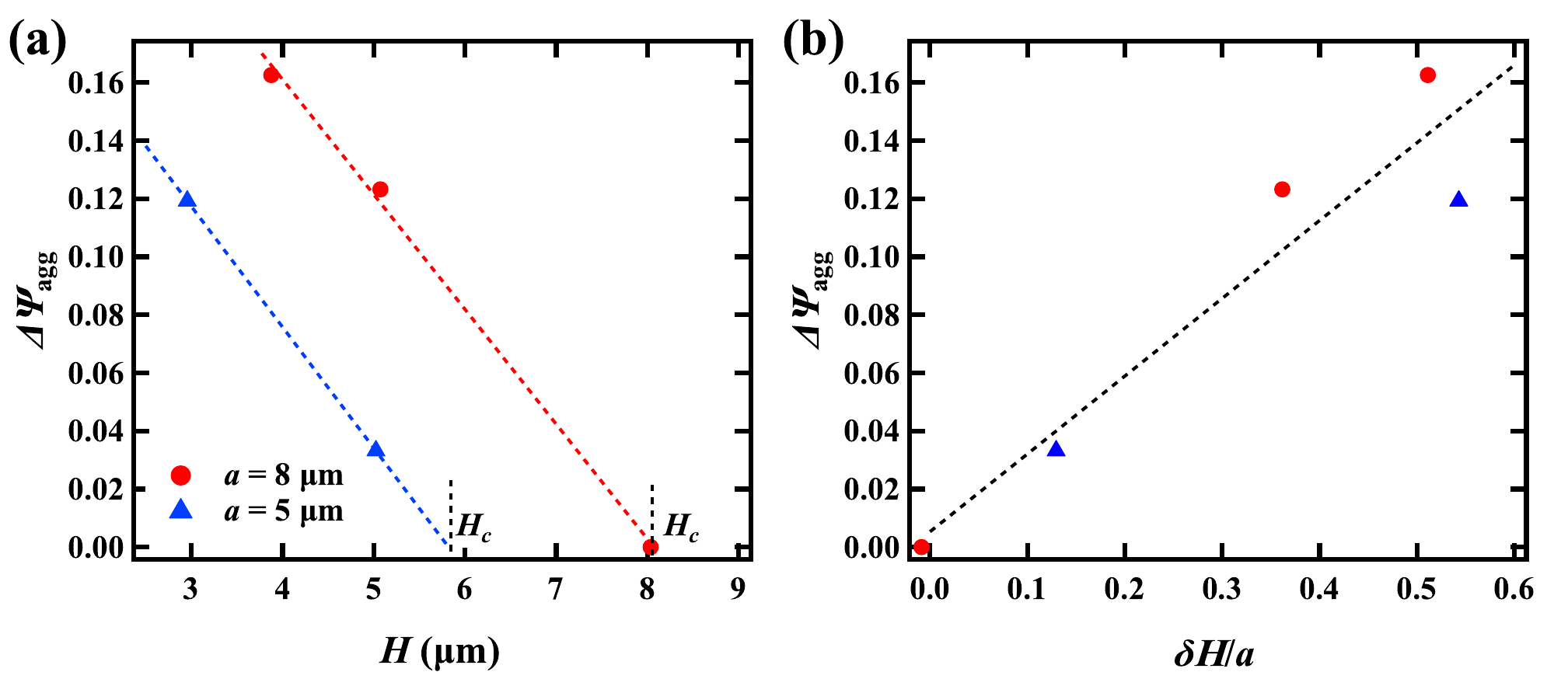}
\caption{\label{fig:dPsi}
(a) Dependence of $\Delta \Psi_{\rm agg}$ on the mean interparticle gap $H$. 
$H_c$ denotes the critical gap at which $\Delta \Psi_{\rm agg}$ becomes approximately zero. 
(b) $\Delta \Psi_{\rm agg}$ plotted as a function of the reduced gap $\delta H/a$. 
Data obtained for different particle sizes collapse onto a single linear relationship. 
This scaling behavior is consistent with a simple geometric rescaling argument, in which the probability distribution of interparticle gaps follows a common form when normalized by particle size.
}
\end{figure}

Figure~\ref{fig:dPsi}(a) shows the dependence of $\Delta \Psi_{\rm agg}$ on the mean interparticle gap $H$. 
For all particle sizes, $\Delta \Psi_{\rm agg}$ decreases monotonically with increasing $H$, exhibiting nearly parallel trends with particle-size-dependent offsets. 
The characteristic scale of $H$ at which $\Delta \Psi_{\rm agg}$ becomes finite is on the order of the particle size $a$. 
This length scale is significantly larger than the polymer radius of gyration that governs classical depletion interactions, indicating that the effective attraction during curing acts over a distinct and much longer range. 
Such a spatial range is consistent with the elasticity-mediated interaction predicted by simulations~\cite{Furuta2024-zt, Furuta2025-vu}. 
We therefore attribute the observed $H$ dependence to the finite spatial range of the effective attraction emerging during curing.

We define $H_c$ as the mean interparticle gap at which $\Delta \Psi_{\rm agg}$ becomes approximately zero within experimental uncertainty. 
From Fig.~\ref{fig:dPsi}(a), we obtain $H_c = 5.8\,\rm{\mu m}$ and $8.0\,\rm{\mu m}$ for $a = 5\,\rm{\mu m}$ and $8\,\rm{\mu m}$, respectively. 
Since $H_c$ appears to scale proportionally with $a$, we introduce a dimensionless reduced gap defined as
\[
\delta H / a \equiv (H_c - H)/a,
\]
and examine its correlation with $\Delta \Psi_{\rm agg}$. 
Figure~\ref{fig:dPsi}(b) shows $\Delta \Psi_{\rm agg}$ as a function of $\delta H/a$, where the data collapse onto a single linear relationship. 
The physical origin of this scaling will be discussed in the following section.

Because $H_c \sim a$, a simple approximation yields a critical volume fraction of $\varphi_c \approx \pi/48 \approx 0.065$. 
For volume fractions exceeding this value, aggregation is promoted during curing under the present experimental conditions.

\subsection{\label{sec:level3E}Exclusion of conventional mechanisms of filler aggregation}

We examined whether the aggregation observed during curing can be explained by conventional van der Waals (vdW) interactions. 
Owing to the high viscosity and nonpolar nature of the epoxy-amine system, the filler particles behave as non-Brownian particles with negligible electrostatic interactions. 
We therefore estimate the characteristic time scale for vdW-driven aggregation by considering the balance between vdW attraction and viscous drag under overdamped conditions.

The equation of motion is given by
\begin{eqnarray}
0 = \frac{A_H a}{24 h^2} - 3 \pi \eta a v,
\end{eqnarray}
where $A_H$ is the Hamaker constant, $\eta$ is the viscosity, and $v$ is the particle velocity. 
From this expression, the characteristic contact time $\tau_{\rm contact}$ can be estimated as
\begin{eqnarray}
\tau_{\rm contact} = \frac{12 \pi \eta}{A_H} a^3 \left( \sqrt[3]{\frac{\pi}{6\varphi}} - 1 \right).
\end{eqnarray}

Because the viscosity increases rapidly during curing, aggregation, if driven solely by vdW forces, would need to occur within the initial 2-3 hours. 
Subsequently, the viscosity exceeds $40~\mathrm{Pa \cdot s}$, as measured by rheometry, and particle motion becomes effectively arrested. 
We therefore determine the critical volume fraction $\varphi_c$ for which $\tau_{\rm contact}$ equals 3 hours. 
Substituting representative parameters ($A_H = 10^{-20}~\mathrm{J}$, $\eta = 0.4~\mathrm{Pa \cdot s}$ before curing, and $a = 5~\mathrm{\mu m}$), we obtain $\varphi_c \approx 0.32$. 
Since this estimate neglects the increase in viscosity during $\tau_{\rm contact}$, the actual critical volume fraction would be even higher. 
This analysis indicates that, under the present experimental conditions, vdW interactions alone are insufficient to induce aggregation at the investigated volume fractions.

\begin{figure}[htbp]
\includegraphics[width=75mm]{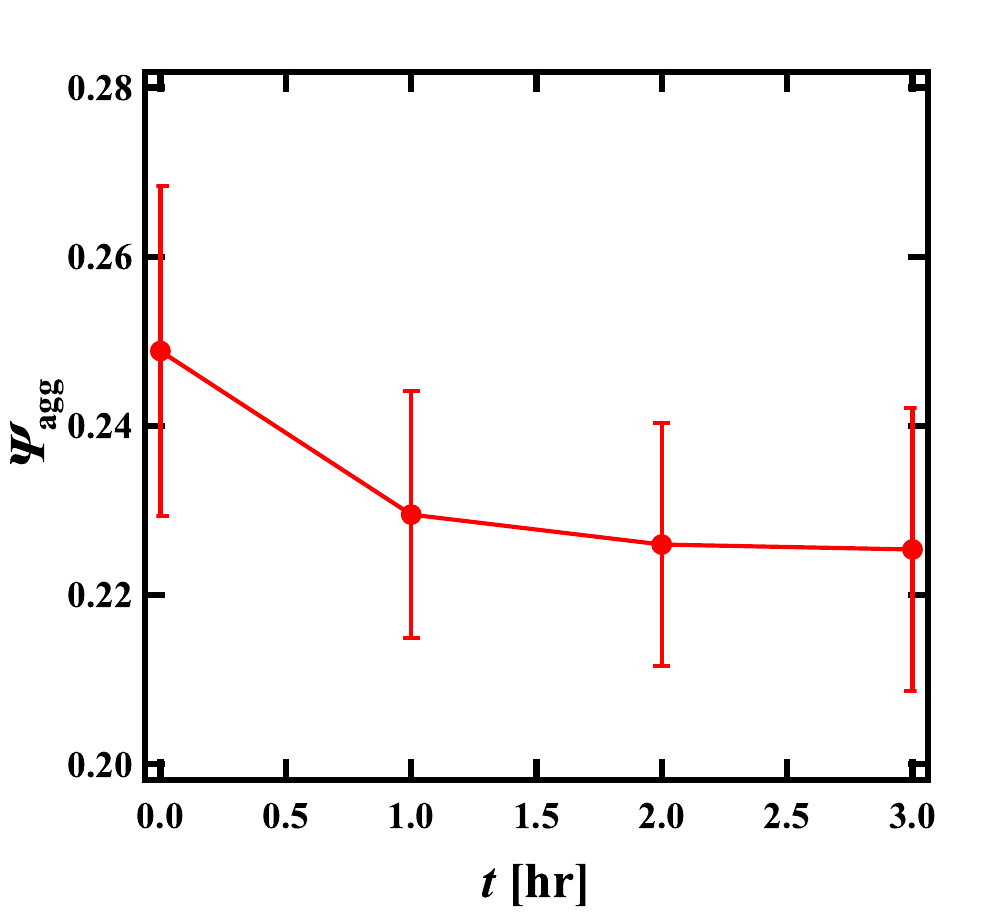}
\caption{\label{fig:epo}
Temporal evolution of the aggregation fraction $\Psi_{\rm agg}$ in the epoxy-FPS system without curing agent at $a = 5\,\rm{\mu m}$ and $\varphi = 0.065$. 
No significant change is observed over time.
}
\end{figure}

To further verify this conclusion, we measured the temporal evolution of the aggregation fraction $\Psi_{\rm agg}$ in a control system consisting of epoxy resin and fluorescent polystyrene beads without the amine curing agent. 
The results are shown in Fig.~\ref{fig:epo}. 
No increase in $\Psi_{\rm agg}$ is observed over time, confirming that vdW interactions alone do not lead to measurable aggregation. 
Although density mismatch between the particles and matrix could, in principle, induce aggregation through differential buoyant velocities~\cite{Turetta2022-vg}, the experimental results demonstrate that such buoyancy-driven effects are negligible under the present conditions. 

Taken together, these findings indicate that the aggregation observed during curing cannot be attributed to conventional vdW forces or buoyancy effects, but instead arises from mechanisms intrinsic to the curing process itself.

\section{\label{sec:level4}Discussions}
We now discuss the physical origin of the observed scaling $\Delta \Psi_{\rm agg} \propto \delta H/a$. 
To this end, we distinguish between two characteristic length scales: 
(i) the intrinsic interaction range $\xi$ associated with the curing-induced effective attraction, and 
(ii) the critical mean interparticle gap $H_c$ defined experimentally as the value of $H$ at which $\Delta \Psi_{\rm agg}$ becomes negligible.
We assume that the surface-to-surface interparticle gap $h$ follows a probability distribution function $P(h; H, a)$ centered at the mean gap $H$. 
Clusters already aggregated at the initial stage are excluded from this consideration. 
Aggregation occurs when a finite fraction of particle pairs satisfy $h \lesssim \xi$, where $\xi$ represents the maximum separation over which the curing-induced effective attraction can promote aggregation. 
Simulation studies suggest that this interaction range scales with particle size, i.e., $\xi \sim a$, and that the potential depth also increases approximately proportionally to $a$~\cite{Furuta2024-zt}.
The aggregation fraction can be expressed as
\[
\Psi_{\rm agg}(H) \propto \int_0^{\xi} P(h; H, a)\, dh.
\]
The condition $H = H_c$ corresponds to the situation in which the probability weight within $h \le \xi$ becomes negligibly small:
\[
\Psi_{\rm agg}(H_c) \approx 0.
\]
Importantly, $H_c$ is not identical to $\xi$; rather, it reflects the combined effect of the intrinsic interaction range $\xi$ and the statistical distribution of interparticle gaps.
When the mean gap decreases to $H = H_c - \delta H$, we approximate this change as a parallel shift of the distribution:
\[
P(h; H_c - \delta H, a) \approx P(h + \delta H; H_c, a).
\]
The aggregation increment is then
\[
\Delta \Psi_{\rm agg}
= \int_{\xi}^{\xi + \delta H} P(h; H_c, a)\, dh
\approx P(\xi; H_c, a)\, \delta H,
\]
where the last approximation holds for sufficiently small $\delta H$.
We further assume that the shape of the gap distribution is self-similar with respect to particle size. 
Since $P(h; H_c, a)$ is a probability density function, dimensional consistency requires
\[
P(h; H_c, a) = \frac{1}{a} f\!\left( \frac{h}{a}; \frac{H_c}{a} \right),
\]
where $f$ is a dimensionless function.
Because both the interaction range $\xi$ and the critical gap $H_c$ scale with particle size ($\xi \sim a$ and $H_c \sim a$), the dimensionless quantities $\xi/a$ and $H_c/a$ are approximately constant. 
Substituting this scaling form into the expression for $\Delta \Psi_{\rm agg}$ yields
\[
\Delta \Psi_{\rm agg} \propto \frac{\delta H}{a},
\]
which explains the experimental collapse observed when plotting $\Delta \Psi_{\rm agg}$ as a function of $\delta H/a$.
The prefactor $f(\xi/a; H_c/a)$ is likely related to the strength of the curing-induced effective interaction. 
These considerations clarify why the volume fraction alone does not uniquely determine the aggregation-promoting effect. 
Instead, the mean interparticle gap $H$, which directly reflects geometric proximity relative to the interaction range, serves as the relevant control parameter. 
Although simulations suggest that the potential depth increases with particle size~\cite{Furuta2024-zt}, a quantitative experimental determination of this interaction strength remains an important subject for future investigation.

\section{\label{sec:level5}Conclusion}

In this work, we have experimentally clarified the mechanism of filler aggregation during curing in epoxy-amine composites through three-dimensional confocal microscopy and quantitative structural analysis. 
While previous studies reported curing-dependent changes in filler aggregation or electrical conductivity~\cite{Shamurina1994-lk, Wu2019-wz}, the governing physical parameters remained unidentified and direct three-dimensional evidence was lacking.
We demonstrate that curing enhances aggregation even in non-Brownian systems where conventional van der Waals forces are insufficient to induce particle contact within the relevant time scale. 
A quantitative estimation of the vdW-driven contact time, together with control experiments without curing agent, excludes both van der Waals attraction and buoyancy-induced clustering as primary mechanisms.

Importantly, we show that the aggregation-promoting effect cannot be described solely by the filler volume fraction. 
Instead, the mean interparticle gap $H$ emerges as the relevant geometric control parameter, and the aggregation increment scales with the reduced gap $\delta H/a$. 
This scaling provides a predictive framework linking particle size, volume fraction, and curing-induced structural evolution.
Recent simulation studies proposed that rigidity percolation during curing generates elasticity-mediated effective attractions whose range and strength scale with particle size~\cite{Furuta2024-zt, Furuta2025-vu}. 
The present results provide direct experimental evidence consistent with this scenario, thereby bridging simulation-based predictions and real composite systems. By identifying the governing geometric parameter and establishing a quantitative scaling relation, this work advances the understanding of structure formation during curing and offers practical guidance for controlling filler dispersion in polymer composite materials.

\vspace{1 cm}

\emph{Acknowledgements}
R. K. was supported by JSPS KAKENHI (20H01874). 

\emph{Authors contributions}
RK conceived the project. YF performed the experiments. All authors wrote the manuscript.

\emph{Competing interests} 
The authors declare that they have no competing financial interests. 

\emph{Correspondence} 
Correspondence and requests for materials should be addressed to YF (furuta@gel.t.u-tokyo.ac.jp) and RK (kurita@tmu.ac.jp).

\emph{Availability of Data and Materials}
All data generated or analyzed during this study are included in this published article. 

\bibliography{myref}

\end{document}